# Fractal Behaviour in the $O(3)$ Model

C. Christou, F.K. Diakonos, H. Panagopoulos
Department of Natural Sciences, University of Cyprus,
P.O. Box 537, CY-1678 Nicosia, Cyprus

**Abstract**

We study domain formation in the two-dimensional $O(3)$ model near criticality. The fractal dimension of these domains is determined with good statistical accuracy.

# 1   Introduction

The $D = 2$ nonlinear $O(3)$ sigma model is known to be a prototype model for asymptotically free theories like the $D = 4$ non-abelian gauge theories. In previous studies of $SU(2)$ lattice gauge theory [1] it has been shown that the domains of deconfinement (clusters with vanishing values of the Polyakov loop) have, at the critical temperature, a fractal structure with dimension $D_F \approx 2.5$. Given the close similarities between the $SU(2)$ gauge theory and the $O(3)$ nonlinear sigma model, one can naturally ask whether such fractal structures can also emerge in the $O(3)$ model. There are however some features of the $O(3)$ lattice theory which make the search for such fractal domains a difficult task :

1) There is no phase transition at finite temperature. One can of course consider $T = 0$ as the critical temperature because the correlation length diverges as $T$ tends to zero. On the other hand this means that the order parameter will vanish at any finite temperature. A similar situation occurs in the $1 - d$ Ising model. In that case, a study of intermittency indices [2], possibly connected with an underlying fractal geometry, shows that even in the presence of an external field, where the phase transition occurs at finite temperature, there are no self-similar structures with fractal geometry surviving in the thermodynamic limit.

2) As $T$ tends to zero, one observes large domains each having a uniform orientation of the spins and the system near the critical point tends to magnetize spontaneously. The underlying rotational symmetry of the model is expressed however through long-wavelength fluctuations corresponding to a slowly rotating magnetization as well as short-wavelength fluctuations around this slow evolution. It is the presence of these fluctuations which hinders the definition of the ordered domains.

Using the so-called Symanzik tree improved action we have studied on the lattice the formation of domains with spontaneous magnetization near the critical temperature. We consider here the situation where the lattice size $L$ is of the order of few correlation lengths. A suitable order parameter is introduced and the dimension of domains with non-vanishing value of the order parameter is determined. It turns out that the geometry of these domains is described by a fractal dimension $D_F \approx 1.75$ which is configuration independent.

The paper is organized as follows : In section 2 we give the action of the model considered and the definitions of the observables used in the subsequent calculations. In section 3 we describe the algorithm used for the determination of the observables and of the fractal dimension. Finally in section 4 we present our numerical results and we discuss their consequences.

# 2   Lattice formulation of the model

We use the lattice action :

$$S_{sym} = -\frac{\beta}{2} \sum_{n,\mu} \{\frac{4}{3}\phi_n^a \phi_{n+\mu}^a - \frac{1}{12}\phi_n^a \phi_{n+2\mu}^a\} \quad (1)$$

where $\phi^a$ is a 3-component scalar field subject to the nonlinear constraint : $\phi^a \phi^a = 1$ at each lattice site. The model is a special combination of a nearest and a next-to-nearest neighbour action chosen so that the short distance (ultraviolet) behaviour is markedly more similar to the behaviour of the continuum action [3]. The correlation length $\xi$ defined through the mass gap $ma$ [4] : $\xi = 1/ma$ diverges exponentially as the temperature tends to zero. Near the critical point, configurations with domains of non vanishing magnetization occur with radii varying from the lattice spacing up to the magnitude of the correlation length. The mean magnetization in each independent configuration has a different orientation. Due to the rotational symmetry of the action configurations with different orientation of the mean magnetization are equally probable. We characterize each configuration by the magnitude $M$ of its mean magnetization:

$$M^a = \frac{1}{L^2} \sum_n \phi_n^a, \qquad M = \sqrt{M^a M^a} \qquad (2)$$

After equilibrium, $M$ is statistically distinct from that of random configurations; the latter typically lead to values of $M$ smaller by 2 orders of magnitude (see Fig.1). This suggests the introduction of the constraint effective potential (CEP) $U_{eff}$ [5] which is a function of $M$ only and is defined by rewriting the finite lattice path integral in the form :

$$Z = \int dM M^2 e^{-U_{eff}(M)} \qquad (3)$$

The free energy differences for various configurations of the theory are proportional to the difference of the corresponding values of the CEP only. The form of CEP is similar to that of the effective potential of a theory undergoing a second-order phase transition. We conjecture that this property is the indication of the self-similarity in the geometry of the ordered domains which in turn leads to the fractality of the system.

Given a configuration with spontaneous mean magnetization of magnitude $M$, its domains are determined by evaluating the quantity :

$$P = \theta(\frac{\phi_n^a M^a}{M} - \epsilon) \qquad (4)$$

$P$ plays the role of the order parameter in our considerations. We introduce also a cut-off parameter $\epsilon$ in order to take into account the precision of the spins due to small wavelength fluctuations. Varying $\epsilon$ we can estimate the effect of these fluctuations to the calculated fractal dimension. If an underlying fractal geometry is present we expect $D_F$ to depend very weakly on $\epsilon$.

## 3  Determining the fractal dimension

After equilibrium is achieved we use Eq.(4) to determine the ordered clusters for each independent configuration. The resulting geometry of the clusters for some configurations is shown in Fig.2. For each configuration we calculate the fractal dimension of the ordered domains. We define the diameter of each domain $D$ through

$$d_D = \max |\mathbf{x_i} - \mathbf{x_j}|, \qquad \forall x_i, x_j \in D \qquad (5)$$

and we include in our calculation only domains greater than $d_{\min}$, where $d_{\min}$ is determined as a fraction of the correlation length $\xi$, such that a) the results are not affected by the presence of pointlike domains and b) we consider clusters of many different sizes. In actual runs we took $d_{\min} = \xi/15$. The fractal dimension is determined using the correlation function defined in [6]:

$$C(R) = \frac{1}{N^2} \sum_{i,j=1}^{N} \theta(R - |\mathbf{x_i} - \mathbf{x_j}|) \qquad (6)$$

where the summation extends over all points with $P = 1$ (cf. Eq.(4)). For a wide range of values of $R$ one expects:

$$C(R) = \alpha R^{D_F} \qquad (7)$$

Indeed, our data exhibit this behaviour, for $d_{\min} \lesssim R \leq L/2$. A typical sample is shown in Fig. 3. Fitting our data to the power law (7) we extract $D_F$ with a statistical error coming from an average over configurations. It is clear from the figure that changing $\epsilon$ has a negligible effect on $D_F$. The calculated fractal dimension shows a weak dependence on the configuration.

It is interesting to study the dependence of the fractal dimension on the temperature. By performing the calculation of $D_F$ at a higher $\beta$ we must take into account the scaling properties of the system. Therefore, due to the increase of $\xi$, which is the only existing physical length scale, we must renormalize the value of $\epsilon$ so that the domains which we consider for determining $D_F$ remain of the same size in units of $\xi$. One way to do this, is to calculate the mean diameter of the domains for all configurations, defined by:

$$\bar{d} = \left\langle \frac{\sum_D N_D d_D}{\sum_D N_D} \right\rangle_C \qquad (8)$$

where the summation is over the domains $D$ and the configurations $C$. $N_D$ is the number of points in each such domain. We then determine $\epsilon$ by requiring:

$$\frac{\bar{d}}{\xi}\Big|_\beta = \frac{\bar{d}}{\xi}\Big|_{\beta'} \qquad (9)$$

Notice that the size of the domains is a function of $\epsilon$. We find that the geometrical picture produced with the adjusted value of $\epsilon$ leads us to approximately the same value of $D_F$.

## 4 Numerical results and conclusions

Our production runs were carried out at 3 different values of $\beta$, listed in Table I. We used the "overheat bath" updating algorithm [7] to thermalize our configurations. The exponential autocorrelation time $\tau_{exp}$ [8] was found to be $< 1000$ for all $\beta$'s used. After a few thousand thermalization sweeps we used a mixture of microcanonical(80%)-overheat bath(20%) algorithms for our measurements. Table I also presents the autocorrelation times $\tau_{int}$, using the mixed algorithm, for the various $\beta$-values. For each $\beta$ we produced an ensemble of 50 independent thermalized configurations, measuring every 50 sweeps for the first two $\beta$-values and every 100 sweeps for $\beta = 1.68$, which was sufficient to determine $D_F$ with a statistical error smaller than 1%.

For any finite value of $\beta$, one may expect to observe fractal behaviour at best up to some maximum length scale: So long as $\xi$ is finite, it is easy to see that, for $R \gg \xi$, definition (6) will lead to $C(R) \propto R^{D_e}$, where $D_e$ is the embedding dimension ($D_e = 2$ in our case). Thus, further data (at larger $R$) in Fig.3 will necessarily tend toward a line with slope 2. To find the range of length scales with fractal behaviour, one must determine the greatest interval of values for $R$, taken around $\xi$, such that the corresponding data in Fig.3 can be fitted by a straight line at a certain confidence level. Clearly, lattice sizes $L$ greater than this maximum range are unnecessary. We chose $L$ so that $L/\xi$ is a constant, for uniformity in comparing results at different $\beta$. We took $L/\xi = 5$; the corresponding values are listed in Table I. With this choice of $L$, fractality was seen to extend out to at least $L/2$. The values of $\xi$ given in Table I are taken from [4]. Any error in the measurement of $\xi$ can influence the precision in the estimation of $\bar{d}/\xi$, but has little effect on the measured value of the fractal dimension. To establish a fractal dimension in this system it is imperative to verify its independence from $\epsilon$, in view of the fact that domain size is $\epsilon$-dependent. The allowed range of values for $\epsilon$ is dictated by the average domain size, which must be neither pointlike nor comparable to the lattice itself. We verified that, within this range, $D_F$ is indeed $\epsilon$-independent. We list in Table I the results for two sets of values for $\epsilon$, chosen in a way as to satisfy Eq.(9) at different $\beta$. The errors reported in Table I are purely statistical, while a comparison at different $\beta$ gives an estimate of possible systematic errors.

The study presented in this paper can be taken over to simulations of QCD with different fermionic content. These afford us with a rich spectrum of phase transitions, some of which could show fractal behaviour. It would also be interesting to compare the fractal dimension in these cases to the critical indices coming from intermittency, as observed in heavy ion collisions. We expect to return to this subject in the future.

**Acknowledgements.** We would like to thank N.G. Antoniou for a number of useful conversations.

# Figure captions

**Fig. 1** : Distribution of $M$ over different configurations at $\beta = 1.68$. The solid spike is the distribution corresponding to random configurations.

**Fig. 2** : A typical pattern of domains. ($\beta = 1.68$, $\epsilon = 0.8$)

**Fig. 3** : $C(R)$ vs. $R$ for different values of $\epsilon$.

# Table 1

| $\beta$ | $\tau_{int}$ (sweeps) | $\xi$ | $L$ | $\epsilon$ | $\bar{d}/\xi$ | $D_F$ |
|---|---|---|---|---|---|---|
| 1.44 | 24 | $19.9 \pm 0.2$ | 99 | 0.850 | $0.198 \pm 0.006$ | $1.742 \pm 0.011$ |
|  |  |  |  | 0.900 | $0.142 \pm 0.003$ | $1.740 \pm 0.018$ |
| 1.56 | 38 | $41.7 \pm 0.3$ | 210 | 0.790 | $0.195 \pm 0.004$ | $1.760 \pm 0.012$ |
|  |  |  |  | 0.845 | $0.138 \pm 0.002$ | $1.753 \pm 0.012$ |
| 1.68 | 75 | $83 \pm 2$ | 399 | 0.750 | $0.195 \pm 0.006$ | $1.779 \pm 0.010$ |
|  |  |  |  | 0.800 | $0.139 \pm 0.004$ | $1.785 \pm 0.014$ |

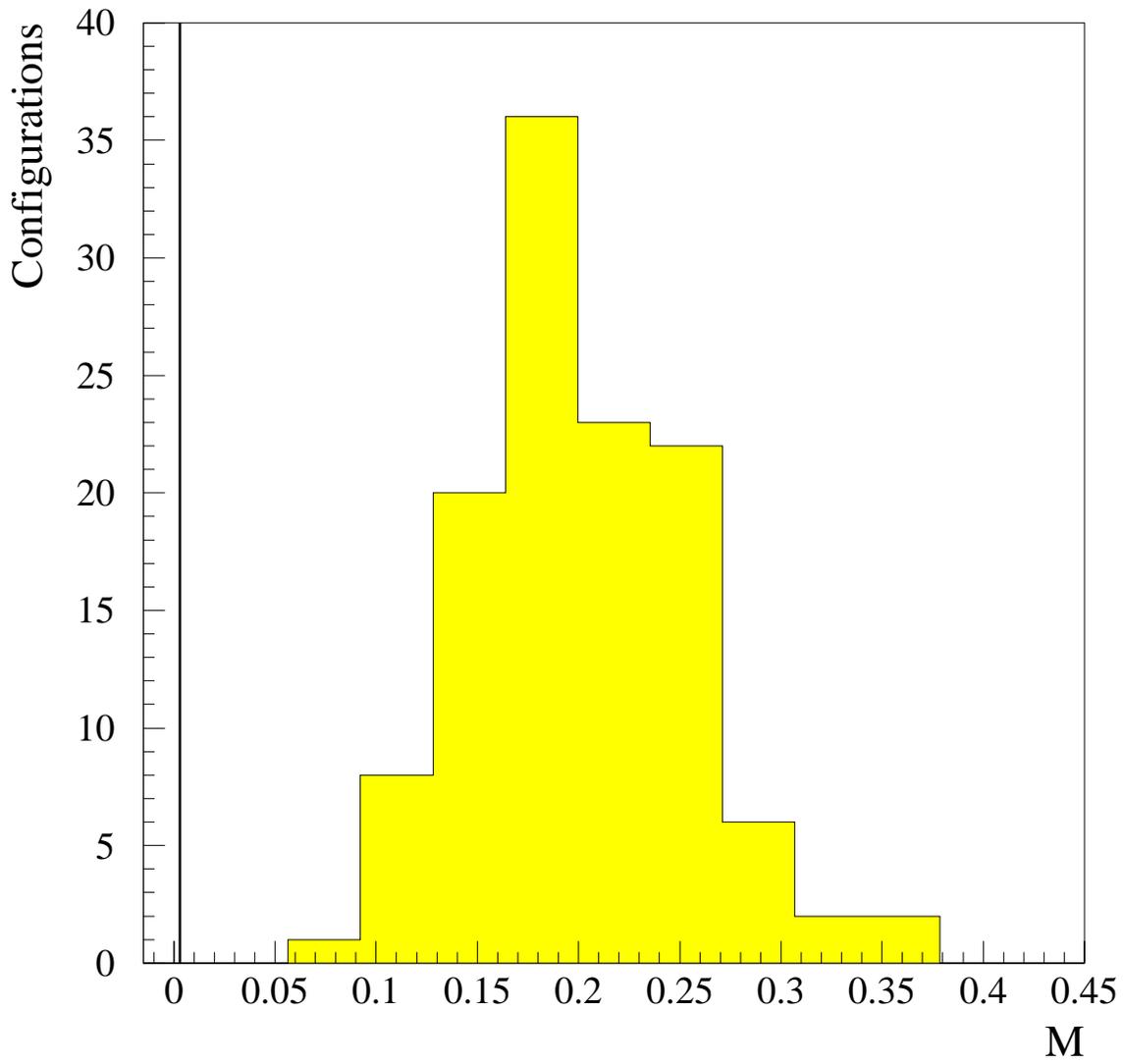

Figure 1

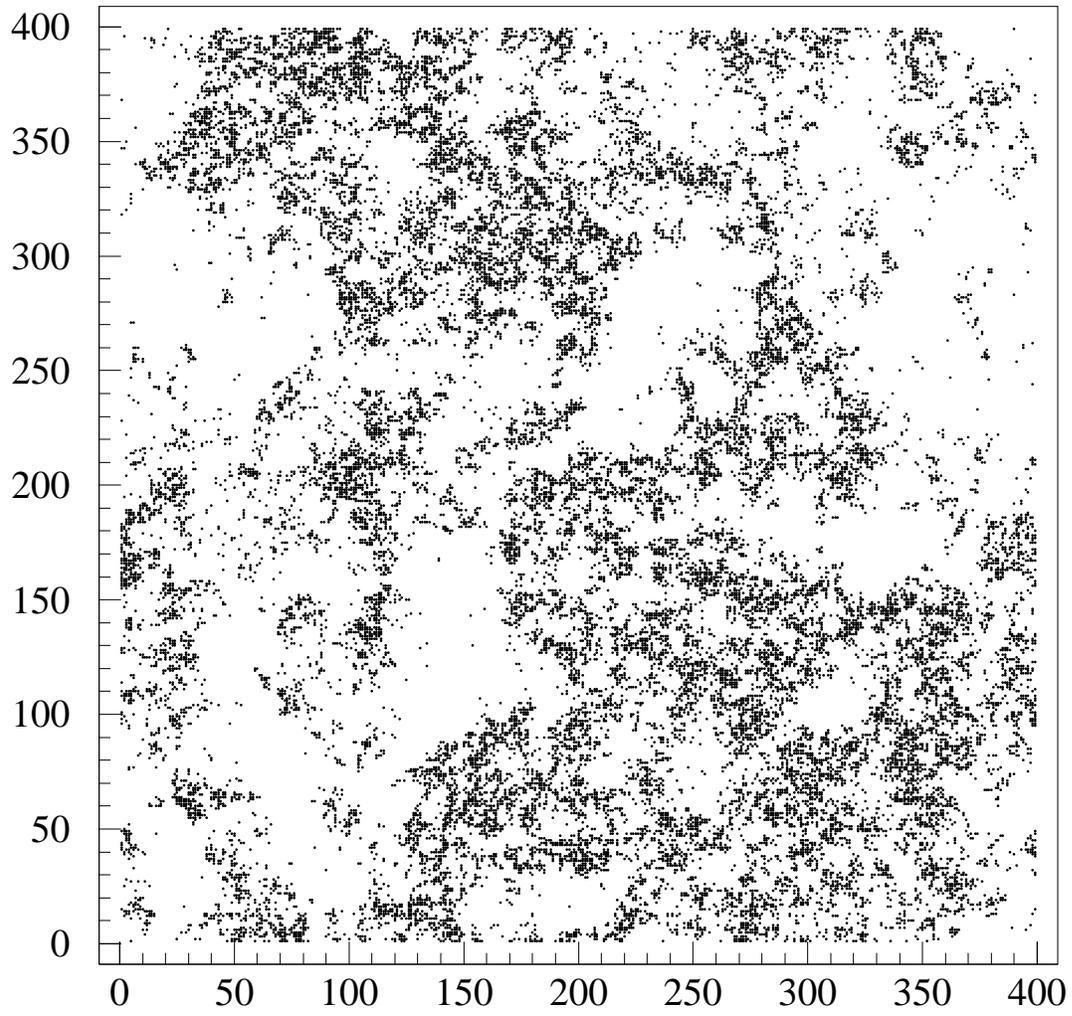

Figure 2

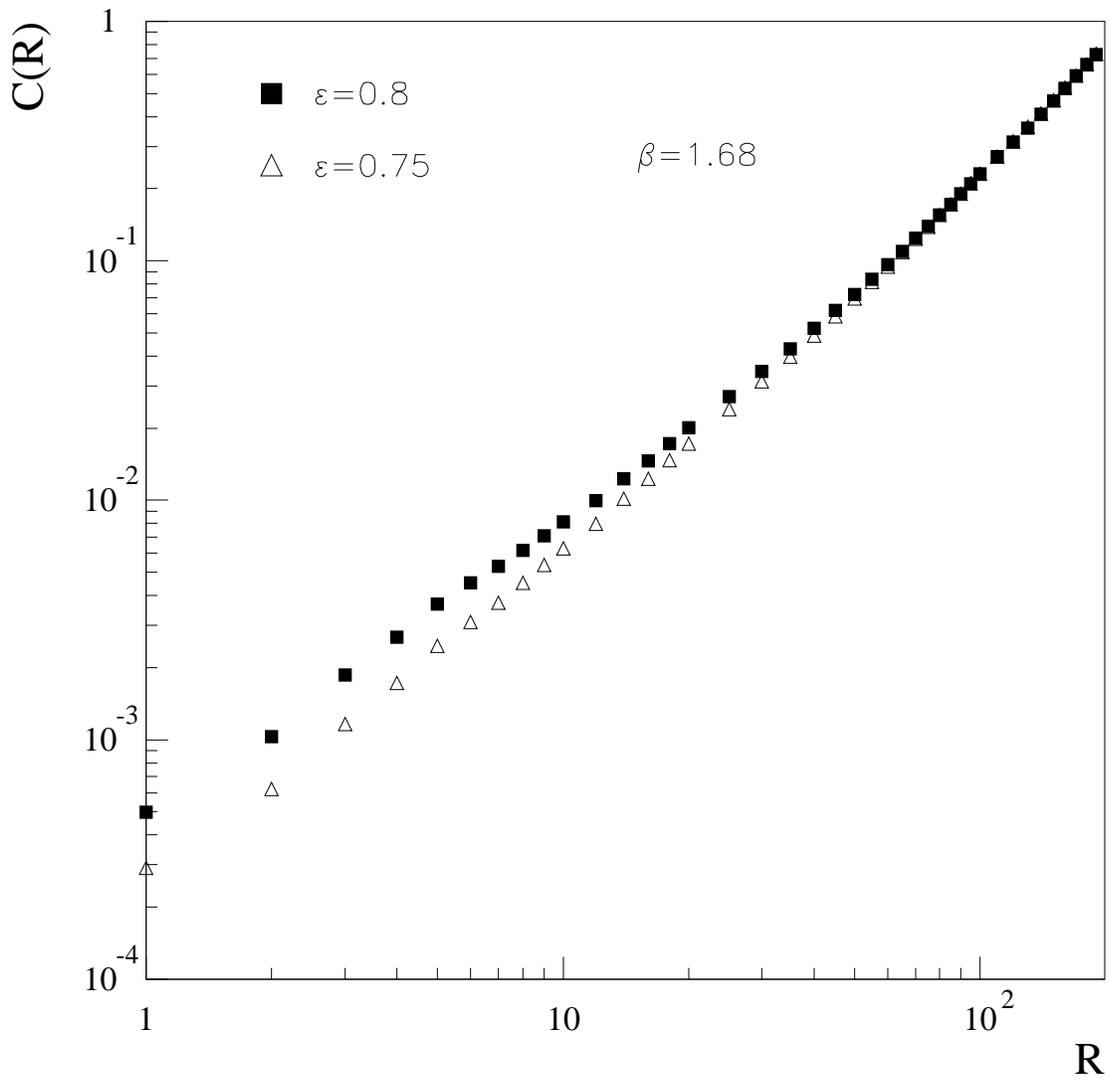

Figure 3